\documentclass[journal=jacsat,manuscript=article]{achemso}

\usepackage[T1]{fontenc} 

\usepackage[version=3]{mhchem}
\usepackage{multirow}

\author{Markus Schneider}
\affiliation[FHI Berlin]
{Fritz-Haber-Institut der Max-Planck-Gesellschaft, Theory Department, Faradayweg 4-6, D-14195 Berlin, Germany}
\author{Carsten Baldauf}
\affiliation[FHI Berlin]
{Fritz-Haber-Institut der Max-Planck-Gesellschaft, Theory Department, Faradayweg 4-6, D-14195 Berlin, Germany}
\alsoaffiliation[WOI Leipzig]
{Wilhelm-Ostwald-Institut f\"ur Physikalische und Theoretische Chemie, Fakult\"at f\"ur Chemie und Mineralogie, Universit\"at Leipzig, Linn\'{e}strasse 2, 04103 Leipzig, Germany}
\email{baldauf@fhi-berlin.mpg.de}

\title
  {Relative energetics of acetyl-histidine protomers with and without Zn$^{2+}$ and a benchmark of energy methods}

\abbreviations{IR,NMR,UV}

\begin{document}



\newpage

\begin{abstract}
\noindent We studied acetylhistidine (AcH), bare or microsolvated with a zinc cation by simulations in isolation.
First, a global search for minima of the potential energy surface combining both, empirical and first-principles methods, is performed individually for either one of five possible protonation states.
Comparing the most stable structures between tautomeric forms of negatively charged AcH shows a clear preference for conformers with the neutral imidazole ring protonated at the \ce{N_{$\epsilon$2}} atom.
When adding a zinc cation to the system, the situation is reversed and \ce{N_{$\delta$1}}-protonated structures are energetically more favorable. 
Obtained minima structures then served as basis for a benchmark study to examine the goodness of commonly applied levels of theory, \textit{i.e.} force fields, semi-empirical methods, density-functional approximations (DFA), and wavefunction-based methods with respect to high-level coupled-cluster calculations, \textit{i.e.} the DLPNO-CCSD(T) method.
All tested force fields and semi-empirical methods show a poor performance in reproducing the energy hierarchies of conformers, in particular of systems involving the zinc cation.
Meta-GGA, hybrid, double hybrid DFAs, and the MP2 method are able to describe the energetics of the reference method within ``chemical accuracy'', \textit{i.e.} with a mean absolute error of less than $1\,\mathrm{kcal/mol}$.
Best performance is found for the double hybrid DFA B3LYP+XYG3 with a mean absolute error of $0.7\,\mathrm{kcal/mol}$ and a maximum error of $1.8\,\mathrm{kcal/mol}$.
While MP2 performs similarly as B3LYP+XYG3, computational costs, i.e. timings, are increased by a factor of 4 in comparison due to the large basis sets required for accurate results.
\end{abstract}

\newpage

\section{Introduction}
Metal cations are essential to life, as approximately one third of the proteins in the human body require a metal cofactor for biological function~\cite{Metalloproteomics,BiologicalInorganicChemistry}.
They often play a crucial role in shaping the three-dimensional structure of proteins and peptides.
In their presence, peptides often undergo significant conformational changes -~usually imposing structural constraints~- that may alter important properties, \textit{e.g.} binding sites, catalytic properties, and biological functions.
As an example, it is hypothesized that protein misfolding of Alzheimer's A$\mathrm{\beta}$-amyloid peptides into aggregated senile plaques inside the human brain of Alzheimer patients is promoted by metal ions such as zinc (\ce{Zn^{2+}})~\cite{Alzheimer}.
Zinc ions are furthermore required for the catalytic function of more than 200 enzymes~\cite{CatalyticZinc}, an example being carbonic anhydrase essential to the process of carbon dioxide regulation~\cite{CarbonicAnhydrase}.

It is therefore much desirable to have a very good fundamental and detailed theoretical understanding of interactions of metal cations with peptides.
The goal of this research is to investigate the energetics of peptides in conjunction with metal cations, with a strong focus on benchmark systems consisting of either a bare acetylhistidine (AcH) or microsolvated with a \ce{Zn^{2+}} cation.
The goodness of commonly applied theoretical levels of theory, \textit{i.e.} force field (FF), semi-empirical quantum chemistry methods, density-functional approximation (DFA), and wavefunction-based methods is being assessed and evaluated with respect to high-level coupled-cluster calculations.
The main focus lies thereby on peptide structures in the gas-phase as opposed to solvated systems.
This is mainly due to the desire of studying the inner, ``undamped'' interactions and forces of molecules and their correct description by theoretical models without having to deal with models outside of the core problem, \textit{e.g.} implicit solvent treatment models, that provide an additional non-negligible uncertainty.
Furthermore, the chosen system of AcH serves as a benchmark system as it is still computationally feasible, even for high-level methods, yet provides a challenging structure because of the tautomeric form of its neutral imidazole ring.

\begin{figure}
  \includegraphics[width=0.8\textwidth]{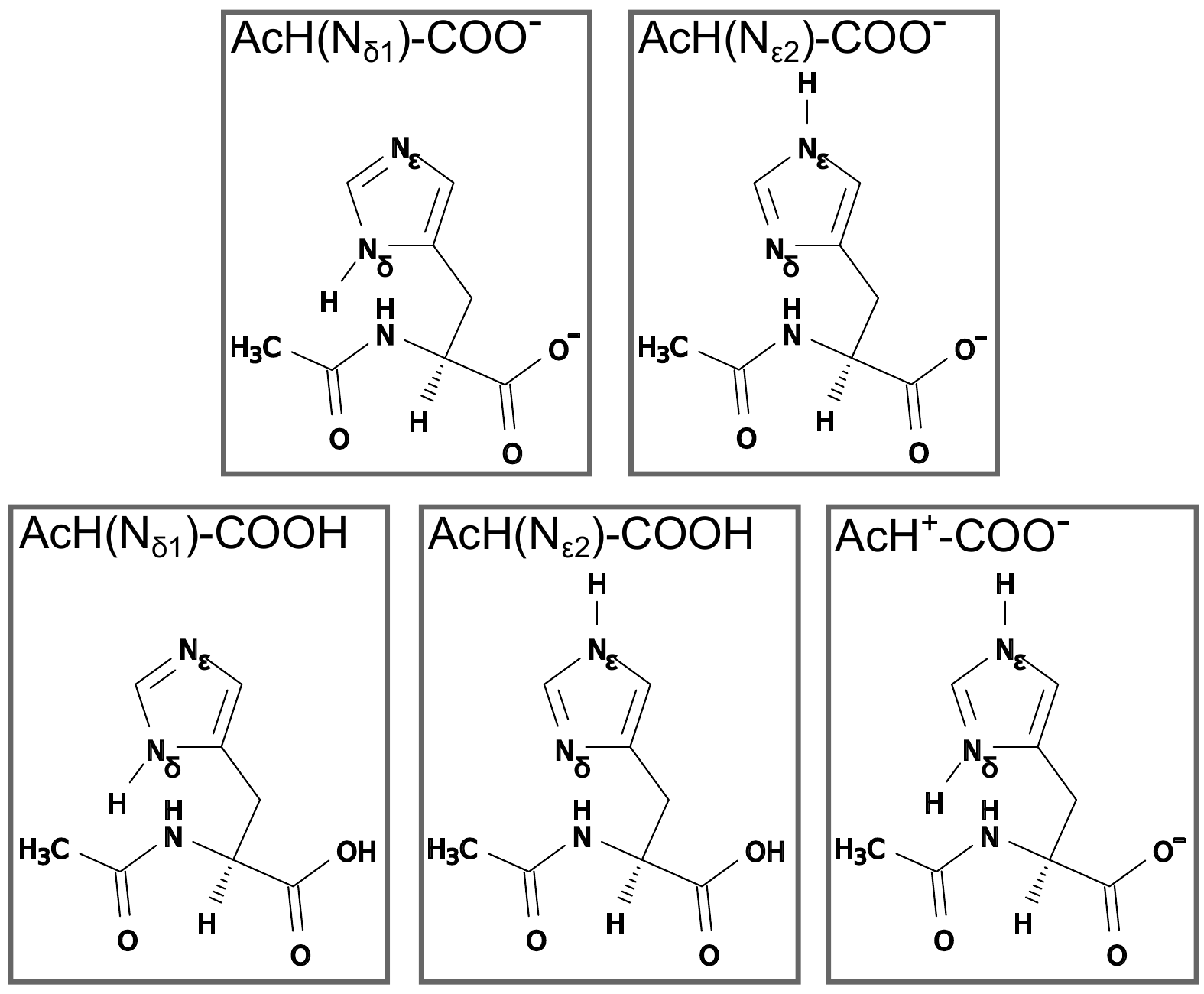}%
  \caption{\label{fig:Tautomers} Chemical structures of negatively charged AcH (upper row) showing the two equivalent tautomeric forms of the neutral imidazole side chain. For neutral AcH (bottom row), three different protonation states are theoretically possible.}
\end{figure}

Figure~\ref{fig:Tautomers} shows chemical structures of AcH with the different protonation states investigated in this work:
Negatively charged AcH (upper row in Figure~\ref{fig:Tautomers}) has two equivalent tautomeric forms of the neutral imidazole side chain.
The two forms are labeled \ce{AcH(N_{$\delta$1})-COO-} and \ce{AcH(N_{$\epsilon$2})-COO-}, meaning that either the \ce{N_{$\delta$1}} or the \ce{N_{$\epsilon$2}} atom is protonated in the imidazole ring.
For bare neutral AcH (bottom row in Figure~\ref{fig:Tautomers}), three different protonation states are theoretically possible:
Besides the two equivalent tautomeric forms, labeled \ce{AcH(N_{$\delta$1})-COOH} and \ce{AcH(N_{$\epsilon$2})-COOH}, that have a neutral carboxyl group at the C-terminus (\ce{-COOH}), a third form exists, labeled \ce{AcH+-COO-}, which has both the \ce{N_{$\delta$1}} and \ce{N_{$\epsilon$2}} nitrogens of the imidazole protonated but the carboxyl group at the C-terminus deprotonated (\ce{-COO^-}).
As already pointed out, either system is studied bare as well as microsolvated with a \ce{Zn^{2+}} cation, resulting in ten different systems to be investigated.

Benchmark calculations for small systems containing a zinc cation have been done in the past. Amin and Truhlar set up a benchmark database of \ce{Zn} coordination compounds with \ce{O}, \ce{S}, \ce{NH3}, \ce{H2O}, \ce{OH}, \ce{SCH3}, and \ce{H} ligands~\cite{ZnBenchmarkLiterature1}.
Using coupled cluster calculations with augmented polarized triple-$\mathrm{\zeta}$ basis sets as the reference, 39 density functionals and seven more approximate molecular orbital theories were tested.
They found that DFT overall significantly outperformed semi-empirical methods.
Best performance was generally found for exchange-correlation functionals containing a portion of Hartree-Fock exchange.
Out of the functionals that contained no Hartree-Fock exchange, M06-L~\cite{M06L} displayed the best performance. 
Similarly, Ray\'{o}n \textit{et al.} tested the performance of five different functionals against MP2~\cite{MP2-1,MP2-2} and CCSD(T)~\cite{CCSDT-3} calculations, with the B3LYP~\cite{B3LYP1,B3LYP2} functional performing best~\cite{ZnBenchmarkLiterature2}.
Weaver \textit{et al.} predicted nine \ce{ZnX} complexes (\ce{X}= \ce{Zn}, \ce{H}, \ce{O}, \ce{F2}, \ce{S}, \ce{Cl}, \ce{Cl2}, \ce{CH3}, \ce{(CH3)2}) using 14 density functionals, MP2 calculations and the CCSD and CCSD(T) coupled-cluster methods applying correlation consistent triple-$\mathrm{\zeta}$ basis sets~\cite{ZnBenchmarkLiterature3}.
They found that BLYP~\cite{BLYP1,BLYP2}, B3LYP, MP2, CCSD and CCSD(T) showed poor performances based on accuracy, which for the latter three wavefunction based methods might be caused by a missing complete basis set description\cite{CBS} or the slow-converging correlation contribution of the zinc electrons that may lead to large and conformation dependent basis set superposition errors (BSSE)~\cite{BSSE-1,BSSE-2}.
Gutten \textit{et al.} evaluated the performance of the wavefunction-based MP2 method as well as several DFA xc functionals with respect to CCSD(T) using gas-phase complexation energies calculated for five model complexes and four metal ions (\ce{Fe^{2+}}, \ce{Cu^{2+}}, \ce{Zn^{2+}}, \ce{Cd^{2+}})~\cite{Rulisek2011}.
Reasonable agreement was found for MP2 with values usually within $1.5\,\mathrm{kcal/mol}$ from the reference values, while DFT performed less satisfactory, although the appropriateness of the models may be significantly altered when combining them with advanced solvation models~\cite{Rulisek20112013}.
For certain complexes containing metal-ligand bonds, large errors in the gas-phase complexation energies (with values up to $20\,\mathrm{kcal/mol}$) were reported.
Performance concerning geometry optimization was found to be satisfactory already using the PBE xc functional on the GGA DFT level.
In the benchmark studies by Navr{\'a}til \textit{et al.} on activation and reaction energies for four model systems of peptide bond hydrolysis in an ion-free environment and in presence of one and two zinc ions, reasonably good performance was found for several DFAs and MP2 when comparing to CCSD(T)-obtained results~\cite{Rulisek2013}.
Best performance for calculating activation barriers was achieved when using the B3LYP or the M06-2X xc functionals on the DFA level of theory.
An extremely large number of amino acid-cation conformations have been generated and analyzed by Ropo and colleagues: The rigorous structure search provided structure/energy data on equal footing over a significant stretch of biochemical space\cite{DataSetRopo} and allowed for interesting analysis, even linking simple structural properties of the amino acid-cation complexes to acute toxicities\cite{AACatTrends}.
Finally, benchmark evaluations and calibrations of theoretical calculations help in modeling metal-binding sites and studying metal-ion selectivity in proteins~\cite{DudevLimReview,DudevLim2001,DudevLim2009,Rulisek2015}.

The general approach in this work follows a previous study by us on the accuracy of various energy functions for carbohydrates \cite{SweetBench} and is briefly outlined in the following.
First, a global search for minima of the potential energy surface (PES) combining both FF and DFA is performed for either one of the ten systems individually.
The obtained global minima and energy hierarchies are then discussed and compared for systems of equal overall charge $q$, \textit{i.e.} $q=1$ for the upper row in Figure~\ref{fig:Tautomers} and $q=0$ for the bottom row in Figure~\ref{fig:Tautomers}.
For the benchmarking studies, a certain set of structures is then selected based on simple energy criteria.
While the focus does lie on local minima structures, it is intended to select structures that vary largely in energy and structure in order to intentionally provide ``difficulty'' for the theoretical methods to be benchmarked.
On top of that, all systems carrying the same overall charge $q$, are benchmarked at once (except for FFs), thus providing even more ``challenge'' for the methods in question.
Finally, across-the-scale total energy calculations for a wide variety of FFs, semi-empirical methods, DFAs, and wavefunction based methods are tested and evaluated against high-level coupled cluster calculations using mean absolute errors (MAEs) and maximum errors (MEs) as a quality measure.

\section{Computational details}
\subsection{Conformational sampling}
In order to yield minima structures that serve as a basis for selecting a set of conformers for the benchmarking process, the conformational space needs to be sampled first.
To that end, an energy minimum search combining both FF and DFA is laid out.
First, a global energy minimum search was performed using a basin-hopping approach within the \texttt{TINKER} molecular modeling package \cite{TINKER1,TINKER2}.
The program works on the level of empirical force fields and the 2009 AMOEBA biopolymer force field, labeled AMOEBA-BIO09~\cite{AMOEBA-1,AMOEBA-2}, was applied here, which is for two reasons:
First, this polarizable force field provides a much ``rougher'' potential energy surface than widely used conventional force fields, such as AMBER-99~\cite{AMBER99}, CHARMM22~\cite{CHARMM22}, or OPLS-AA~\cite{OPLS-AA-1,OPLS-AA-2}, because it uses atomic charge multipole expansion instead of fixed point charges.
The ``rougher'' the potential energy surface the more minima are found, hence the conformational space is sampled in more detail.
For example, depending on the actual studied system, \textit{i.e.} whether the \ce{Zn^{2+}} cation is present or the protonation state of the imidazole side chain and the carboxyl group, the number of minima found can be up to a factor of six higher when using the AMOEBA-BIO09 force field in comparison to the OPLS-AA, AMBER-99, and CHARMM22 force fields.
Secondly, the AMOEBA-BIO09 FF was the only FF available providing out-of-the-box parameters for the neutral carboxyl group (\ce{-COOH}).
Concerning the technical aspect of the basin-hopping search, the \texttt{scan} subprogram within \texttt{TINKER} has been applied using all automatically found torsional angles, a relative energy window of $100\,\mathrm{kcal/mol}$ and an energy similarity criterion of $0.0001\,\mathrm{kcal/mol}$.
After having applied the FF driven basin-hopping approach, all found minima were locally refined, \textit{i.e.} the individual structures were being geometrically relaxed, using DFA implemented within the all-electron/full-potential electronic structure code package \texttt{FHI-aims}~\cite{FHI-aims, FHI-aims-RI, FHI-aims-ELPA}.
To be more precise, relaxation was accomplished using a trust radius method version of the Broyden-Fletcher-Goldfarb-Shanno (BFGS) optimization algorithm \cite{BFGS-TrustRadius}. 
First, local refinement was done on the PBE+vdW level using \texttt{FHI-aims} specific \texttt{tier 1} basis sets and \texttt{light} settings intended to give reliable energies for screening purposes~\cite{FHI-aims}.
This means the PBE~\cite{PBE} generalized gradient approximation (GGA) exchange-correlation (xc) functional has been applied along with the Tkatchenko-Scheffler~\cite{TS} van der Waals scheme (vdW) for treating long-range dispersion effects.
To rule out duplicate structures, a clustering scheme was applied.
To be precise, root-mean-square deviations (RMSD) of atomic positions between any two conformers were calculated using \texttt{OpenBabel}~\cite{OpenBabel}.
Hierarchical clustering was then achieved by applying the Unweighted Pair Group Method with Arithmetic Mean (UPGMA)~\cite{UPGMA} method implemented in \texttt{Python}'s \texttt{SciPy}~\cite{SciPy} library.
Following that, further relaxation was accomplished at the PBE+vdW level using \texttt{tier 2} basis sets and \texttt{tight} settings that are intended to provide $\mathrm{meV}$-level accurate energy differences~\cite{FHI-aims}, \textit{i.e.} within $0.02\,\mathrm{kcal/mol}$.
After clustering, further relaxation was accomplished at the PBE0+MBD level using the same two-step approach as before, \textit{i.e.} using first \texttt{tier 1} basis sets and \texttt{light} settings, and \texttt{tier 2} basis sets and \texttt{tight} settings afterwards.
As the labeling suggests, the PBE0~\cite{PBE0} hybrid xc functional has been applied, augmented by a many-body dispersion (MBD)~\cite{MBD} correction for long-range dispersion treatment.

\subsection{Levels of theory and energy calculation methods}

\subsubsection{Wavefunction-based methods}

The benchmark calculations in this paper are based on high-level coupled-cluster calculations~\cite{CCSDT-1,CCSDT-2}.
In particular, the coupled-cluster method including single, double, and perturbative triple excitations, named CCSD(T)~\cite{CCSDT-3}, is commonly referred to as the ``gold standard of quantum chemistry'' due to its high accuracy in the complete basis set limit (CBS)~\cite{CCSDTaccuracy}.
However, due to the slow convergence of the electronic correlation energy with basis set size $N$ as well as the technique's $\mathcal{O}(N^7)$-scaling of the computational costs, accurate results that require large enough basis sets are currently not affordable for system sizes treated in this work.
Instead, the domain-based local pair natural orbital (DLPNO-)CCSD(T)~\cite{DLPNO1,DLPNO2} technique serves as the reference method in this work.
The DLPNO-CCSD(T) approximation aims to fully exploit locality of the electron correlation and shows a near linear scaling behavior with basis set size $N$.
Calculations have been carried out with the electronic structure program package \texttt{ORCA}\cite{ORCA}.
Ahlrichs' \texttt{def2}\cite{def2} basis set family has been used throughout for all wavefunction-based methods.
Because heavy elements like \ce{Zn^{2+}} require a relativistic treatment of all-electron calculations, the 0\textsuperscript{th} order regular approximation (ZORA)~\cite{ZORA}, implemented in \texttt{ORCA} in an approximate way~\cite{ORCA-ZORA-1}, is used throughout.
As the scalar relativistic treatment requires flexible basis sets, this in turn means that \texttt{ORCA} automatically provides relativistically recontracted versions~\cite{ORCA-ZORA-2} of Ahlrichs' \texttt{def2} basis set family, labeled \texttt{ZORA-def2}.
The accuracy of the DLPNO-CCSD(T) method has been tested previously with a series of benchmark sets covering a broad range of quantum chemical applications~\cite{DLPNOaccuracy}.
An accuracy of $1\,\mathrm{kcal/mol}$ commonly named \textit{chemical accuracy}, could be obtained using \texttt{normal} settings.
Still, before using the DLPNO-CCSD(T) method with \texttt{normal} settings as the reference method in this work, validation has to be done against conventional CCSD(T) calculations for the systems depicted in Figure~\ref{fig:Tautomers} and using Ahlrichs' relativistically recontracted split valence basis set with added polarization functions, labeled ZORA-def2-SVP.

The other post-Hartree-Fock \textit{ab initio} method in this work to be benchmarked against DLPNO-CCSD(T), is the widely used second-order M{\o}ller-Plesset perturbation theory (MP2)~\cite{MP2-1,MP2-2}.
Calculations have been carried out again with \texttt{ORCA} and applying a resolution of identity (RI) approximation~\cite{ORCA-RI}. 

Energy calculations for both DLPNO-CCSD(T) and MP2 have been performed using Ahlrichs's ZORA-def2-SVP basis set as well as relativistically recontracted valence triple-zeta and quadruple-zeta basis sets with two sets of polarization functions added, labeled ZORA-def2-TZVPP and ZORA-def2-QZVPP, respectively.
Extrapolation to the CBS limit has been applied on calculated Hartree-Fock (HF) energies and correlation energies individually.
To be more precise, HF energies have been extrapolated using a form proposed by Karton and Martin~\cite{SCFextrapolation}:
\begin{equation} 
    E^{n}_{HF}=E^{CBS}_{HF}+A e^{-\alpha\sqrt{n}},
    \label{equ:CBSHF}
\end{equation}
with $A$, $\alpha$, and the CBS-extrapolated energy $E^{CBS}_{HF}$ being parameters to be determined from a least-squares fitting algorithm that has been applied individually for each conformer, and $n$ denoting the cardinal number of the respective basis set, \textit{i.e.} $n=2$ for ZORA-def2-SVP, $n=3$ for ZORA-def2-TZVPP, and $n=4$ for ZORA-def2-QZVPP.
A similar extrapolation scheme has also been laid out for the correlation energies, this time following the form proposed by Truhlar~\cite{MP2extrapolation}:     
\begin{equation} 
    E^{n}_{corr}=E^{CBS}_{corr}+B n^{-\beta},
    \label{equ:CBScorr}
\end{equation}
again with $B$, $\beta$, and the CBS-extrapolated energy $E^{CBS}_{corr}$ being parameters to be determined from a least-squares fitting algorithm as before.
Extrapolation using all three basis set families (\textit{i.e.} $n=2,3,4$) has been found to yield inconsistent results between the different systems depicted in Figure~\ref{fig:Tautomers}.
Hence, extrapolation was laid out using only ZORA-def2-TZVPP and ZORA-def2-QZVPP, \textit{i.e.} an effective two-point extrapolation scheme assuming $\beta=3$, as originally proposed by Halkier \textit{et al.}~\cite{ExtrapolationBetaIs3}.

Finally, for systems microsolvated with a \ce{Zn^{2+}} cation, the slow-converging correlation contribution of the zinc electrons may lead to large and conformation dependent basis set superposition errors (BSSE)~\cite{BSSE-1,BSSE-2}.
To account for that and prior to performing CBS extrapolation as described above, we subject the HF and correlation energies of each \ce{Zn^{2+}} coordinated conformation to a counterpoise correction as proposed by Boys and Bernardi~\cite{CP}:
Assuming rigid conformers, the BSSE is estimated as
\begin{align}
  \begin{split}
      E_{\mathrm{BSSE}} = & E_{\mathrm{BSSE}}(\mathrm{AcH}) + E_{\mathrm{BSSE}}(Zn^{2+})\text{ , with}\\
                          & E_{\mathrm{BSSE}}(\mathrm{AcH})     = E^{\mathrm{AcH+Zn^{2+}}}(\mathrm{AcH})     - E^{\mathrm{AcH}}(\mathrm{AcH})\text{ , and}\\
                          & E_{\mathrm{BSSE}}(\mathrm{Zn^{2+}}) = E^{\mathrm{AcH+Zn^{2+}}}(\mathrm{Zn^{2+}}) - E^{\mathrm{Zn^{2+}}}(\mathrm{Zn^{2+}}),
  \end{split}
  \label{equ:BSSE}
\end{align}
where $E^{\mathrm{AcH+Zn^{2+}}}(\mathrm{AcH})$ represents the energy of AcH evaluated in the union of the basis sets on AcH and \ce{Zn^{2+}}, $E^{\mathrm{AcH}}(\mathrm{AcH})$ represents the energy of AcH evaluated in the basis set of AcH, \textit{etc.}
The individual BSSE errors were then subtracted from the HF and correlation energy, respectively.

\subsubsection{Empirical force fields}

Single-point energy calculations using several out-of-the-box force fields (FFs) were carried out using the \texttt{TINKER}~\cite{TINKER2} molecular modeling package.
Two classes of FFs were tackled: (i) conventional FFs, in particular AMBER-99~\cite{AMBER99}, CHARMM22~\cite{CHARMM22}, and OPLS-AA~\cite{OPLS-AA-1,OPLS-AA-2}, as well as (ii) polarizable atomic multipole-based FFs that use atomic charge multipole expansion instead of fixed point charges.
In particular, these are the 2009 AMOEBA biopolymer FF named AMOEBA-BIO09~\cite{AMOEBA-1,AMOEBA-2}, and the 2013 AMOEBA protein FF named AMOEBA-PRO13~\cite{AMOEBA-PRO}.
Note that only for systems containing a deprotonated carboxyl group (\ce{-COO^-}), parameters were available for all force fields out-of-the-box.
As AMOEBA-BIO09 was the only FF available providing also parameters for the neutral carboxyl group (\ce{-COOH}), FF calculations for systems containing neutral AcH (lower row in Figure~\ref{fig:Tautomers}) were only laid out using this particular FF.

\subsubsection{Semi-empirical quantum chemistry methods}

Semi-empirical quantum chemistry methods are based on the Hartree-Fock method, but follow a simplification strategy by making approximations for computationally demanding terms.
In order to account for caused errors, empirical parameters are incorporated into the formalism and fitted against experimental data or high-level calculations~\cite{SQMreview}.
All semi-empirical methods tackled in this work are based on the neglect of diatomic differential overlap (NDDO)~\cite{NDDO}, a method for approximating computational costly three-center and four-center two-electron integrals.
In particular, the different applied models are the Austin Model 1 (AM1)~\cite{AM1}, the Parametric Method 3 (PM3)~\cite{PM3}, the Parametric Method 6 (PM6)~\cite{PM6}, and the Parametric Method 7 (PM7)~\cite{PM7}.
All semi-empirical method calculations have been carried out using the \texttt{MOPAC2016}~\cite{MOPAC2016} semi-empirical quantum chemistry program.
For the specific case of PM6, two additional long-range dispersion correction schemes were tackled as well.
In particular, these are Grimme's D3 correction for dispersion~\cite{D3} plus a simple function for hydrogen bonds, as well as the corrections to hydrogen bonding and dispersion by \v{R}ez\'{a}\v{c} and Hobza~\cite{D3H4-1,D3H4-2}, labeled D3H4.
The corresponding conjunctive methods are then accordingly being labeled PM6-D3 and PM6-D3H4.

While single-point energy calculations carried out for all other methods in this work refer to total energies on the potential-energy surface, semi-empirical methods yield heats of formation. The heat of formation is defined as the sum of the electronic energy, the nuclear-nuclear repulsion energy, the ionization energy for the valence electrons, the total heat of atomization of all the atoms in the system, and -- if available -- the energy from hydrogen bonds and dispersion correction~\cite{HeatOfFormation}.
When comparing potential energies of other computational methods with heats of formation obtained from semi-empirical calculations through the means of MAEs and MEs, the systematic shift between the two is accounted for, as explained below.

\subsubsection{Density-functional approximations}

While density-functional theory (DFT)\cite{DFT1,DFT2} in itself is an exact method, in practice approximations have to be made because the exact form of the exchange-correlation (xc) functional is unknown, except for the free electron gas.
A large variety of different DFAs exist, commonly classified into different types depending on the features and formal properties of the xc functionals in question~\cite{JacobsLadder}.
The ones selected in this work are summarized in the following:
\begin{itemize}
    \item \textit{Generalized gradient approximations} (\textit{GGAs}) are characterized by the dependence of the xc functional only on the electron density and its gradient.
        In this work, we studied the accuracy of the Perdew-Burke-Ernzerhof (PBE)~\cite{PBE} and Becke-Lee-Yang-Parr (BLYP)~\cite{BLYP1,BLYP2} xc functionals.
    \item In addition to GGAs, \textit{meta-GGAs} also depend on the Laplacian of the electron density or include the kinetic energy density.
        We tested the M06-L~\cite{M06L} and M11-L~\cite{M11L} xc functionals from the group of Minnesota functionals, as well as the SCAN~\cite{SCAN} functional.
    \item For the computationally more costly class of \textit{hybrid} functionals, the exchange parts of the functional are admixed with exact exchange from Hartree-Fock theory.
        We tested the PBE0~\cite{PBE0}, B3LYP~\cite{B3LYP1,B3LYP2} and SCAN0~\cite{SCAN0} functionals.
        In addition, several hybrid functionals from the group of Minnesota functionals were tested as well, in particular the M06~\cite{M06}, M06-2X\cite{M06}, M08-SO~\cite{M08etc}, M08-HX~\cite{M08etc}, and M11~\cite{M11} functionals.
\end{itemize}
Calculations for the PBE, BLYP, M11-L, SCAN, PBE0, B3LYP, M08-SO, M08-HX, and M11 xc functionals were carried out with \texttt{FHI-aims} using \texttt{tier 2} basis sets and \texttt{really\_tight} settings, and including a relativistic treatment by applying the atomic ZORA method~\cite{FHI-aims}. The SCAN and SCAN0 functionals are implemented in \texttt{FHI-aims} via the \texttt{dfauto} program~\cite{dfauto}.
Calculations for the M06-L, M06, and M06-2X xc functionals were carried out with \texttt{ORCA}, including ZORA and the relativistically recontracted ZORA-def2-QZVPP basis set, as explained above.

Commonly applied semi-local DFAs and conventional hybrid functionals are unable to capture the essence of long-range dispersion effects~\cite{LongRangeDFT}.
Many systems containing biomolecules rely on \textit{van der Waals} interaction treatments for an accurate energetic description~\cite{DataSetRopo,AACatTrends,going-clean,HelixInIsolation}.
To that end, computationally cheap correction schemes exist that are evaluated \textit{a posteriori}, \textit{i.e.} they are accounted for after the electron density has been obtained via the self-consistent treatment in DFT. Three different vdW correction schemes were tackled in this work:
\begin{itemize}
    \item The general empirical pairwise additive D3 dispersion correction method by Grimme \textit{et al.}~\cite{D3} provides a consistent description across the whole periodic table.
        Here we used the zero-damping function for short ranges, including three-body dispersion contributions.
        In order to match the long- and midrange correlation of D3 with the semilocal correlation computed by the xc functional, the parameterization of the damping function depends on the xc functional itself.
        Hence, only xc functionals where an out-of-the-box D3 treatment was available, were tested.
        In particular, we evaluated M06-L+D3, M06+D3, and M06-2X+D3 using \texttt{ORCA}, using the same settings as described above.
        For the methods of PBE+D3, BLYP+D3, PBE0+D3, and B3LYP+D3, long-range dispersion calculations were done on top of the \texttt{FHI-aims} calculated energies using Grimme's stand-alone program \texttt{DFT-D3}~\cite{DFT-D3}.
    \item The parameter-free pairwise Tkatchenko-Scheffler van der Waals scheme (vdW)~\cite{TS} relies on summing interatomic pairwise, electron-density derived $C_6$ coefficients, and accurate reference data for the free atoms.
        As the method is implemented in \texttt{FHI-aims}, calculations are carried out for the methods of PBE+vdW, BLYP+vdW, PBE0+vdW, and B3LYP+vdW.
    \item In contrast to the previous pairwise Tkatchenko-Scheffler scheme that ignores the intrinsic many-body nature of correlation effects, the many-body dispersion scheme labeled MBD~\cite{MBD} (and sometimes also labeled MBD* or MBD@rsSCS) combines the TS scheme with the self-consistent screening (SCS) equation of classical electrodynamics.
        In addition, a range-separation (rs) technique is applied, separating correlation into a short-range and a long-range contribution.
        Calculations were carried out for the methods of PBE+MBD and PBE0+MBD using \texttt{FHI-aims}.
\end{itemize}

In order to avoid high computational costs of hybrid xc functionals and still yield accurate results, recent focus has been set on ``low-cost'' DFT based \textit{composite electronic structure approaches}.
In particular, the PBEh-3c method by Grimme \textit{et al.}~\cite{PBEh-3c} aims to efficiently compute structures and interaction energies.
It is based on a modified hybrid variant of the PBE GGA xc functional with a relatively large amount of non-local Hartree-Fock-exchange (42\%).
The orbitals are expanded in Ahlrichs-type valence-double zeta atomic orbital Gaussian basis sets.
Furthermore, the D3 scheme and the global counterpoise-correction scheme (gCP)~\cite{gCP} are applied in order to account for long-range dispersion and BSSE effects, respectively.
Calculations were carried out with \texttt{ORCA}.

Finally, \textit{double hybrid xc functionals} extend hybrid xc functionals in a way that both the exchange and the correlation part contain non-local orbital-dependent components. In particular, we test the B3LYP+XYG3~\cite{XYG3} method implemented in \texttt{FHI-aims}.
As the name implies, the exchange parts of the functional are admixed with exact exchange from Hartree-Fock theory, while a fraction of the correlation part is calculated using G\"orling-Levy coupling-constant perturbation expansion to the second order (PT2)~\cite{PT2}.
Calculations were carried out with \texttt{FHI-aims} using numerically tabulated atom-centered orbital triple-zeta basis sets with valence-correlation consistency, labeled NAO-VCC-nZ~\cite{NAO-VCC-nZ}.
Constructed analogous to Dunning's correlation-consistent polarized valence-only basis sets (cc-pVnZ)~\cite{cc-pVnZ}, these basis sets utilize the more flexible shape of NAOs.
Zhang \textit{et al.} showed that XYG3 provides best results in combination with the triple-zeta NAO-VCC-3Z basis set~\cite{NAO-VCC-3Z-best}.
Because the NAO-VCC-3Z basis set was not available out-of-the-box for the element of Zn, we used Dunning's analogous cc-pV3Z~\cite{cc-pV3ZforZn} basis set instead for this particular element.

\subsection{Mean absolute error and maximum error}

\begin{figure}
  \includegraphics[width=0.6\textwidth]{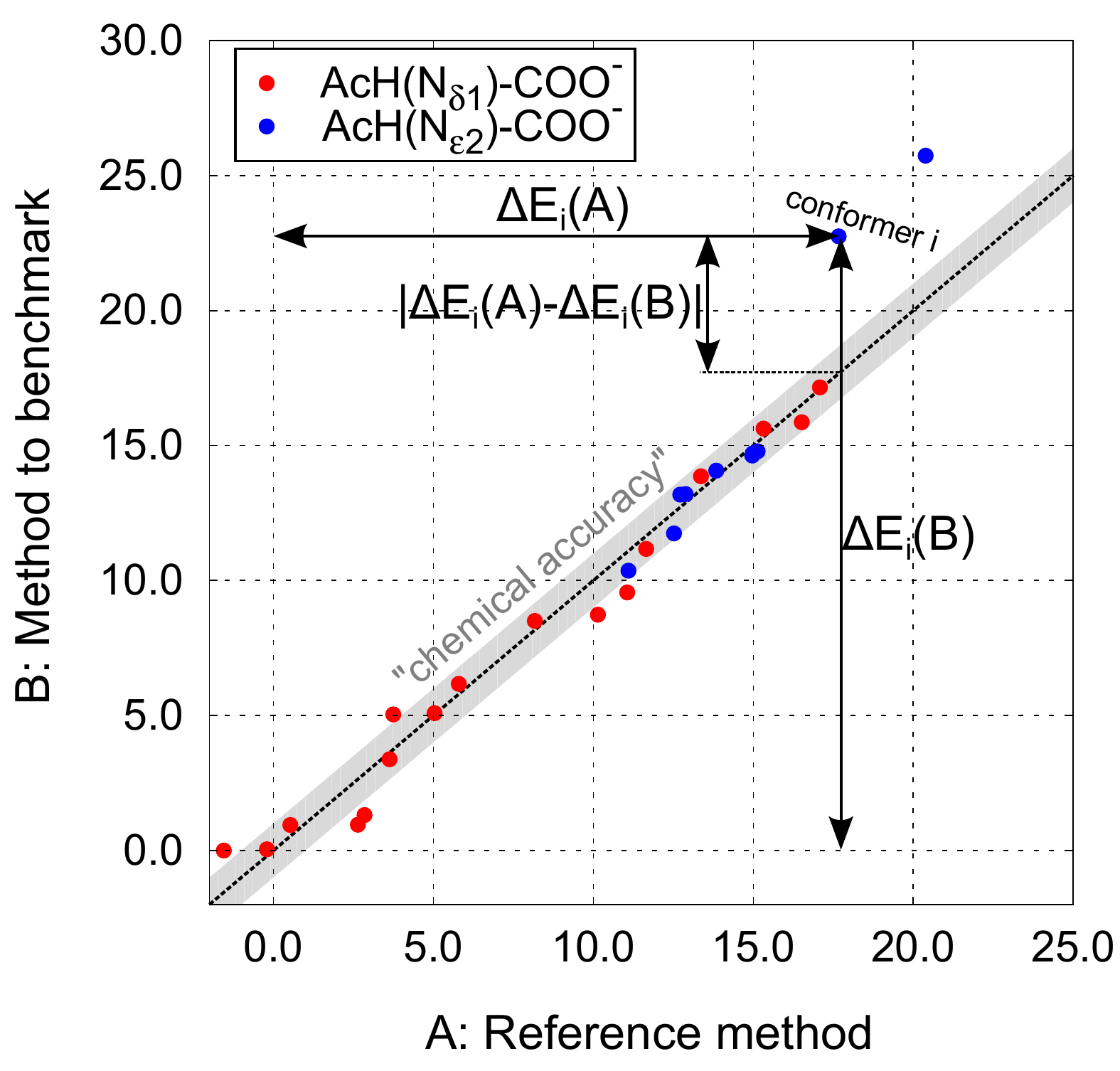}%
  \caption{\label{fig:ScatterExplained} Example of a correlation plot of two different sets of conformers (red and blue). The reference (A) and benchmark (B) conformational hierarchies have already been shifted to minimize the MAE. If the energy description between the reference method and the method to be evaluated agreed perfectly, all points would align on the dashed diagonal line. The gray shading denotes an absolute energy deviation of $1\,\mathrm{kcal/mol}$, \textit{i.e.} the region of ``chemical accuracy''. For a specific conformer $i$, the absolute energy deviation \mbox{$|\Delta E_i^{\mathrm{reference}} - \Delta E_i^{\mathrm{benchmarked}}| = |\Delta E_i(A) - \Delta E_i(B)|$} is illustrated.}
\end{figure}

In order to compare the energetic performance of different methods, single-point energy calculations of a set of different conformers were compared by means of mean absolute errors (MAEs) and maximum errors (MEs).
MAEs of relative energies between the reference method and the method to be benchmarked were calculated as follows:
\begin{equation} 
    MAE = \frac{1}{N} \sum_{i=1}^{N} |\Delta E_i^{\mathrm{reference}} - \Delta E_i^{\mathrm{benchmarked}}+c|,
    \label{equ:MAE}
\end{equation}
where the index $i$ runs over all $N$ conformations of a given data set. 
$\Delta E_i$ in principle denotes the energy difference between conformer $i$ and the lowest-energy conformer of the set. 
The adjustable parameter $c$ is used to systematically shift the reference and benchmark conformational hierarchies versus one another to obtain the lowest possible MAE, rendering the reported MAE value independent of the choice of any reference structure. 
Similarly, MEs were calculated as follows:
\begin{equation} 
    ME = \max_{i\in N} |\Delta E_i^{\mathrm{reference}} - \Delta E_i^{\mathrm{benchmarked}}+c|,
    \label{equ:ME}
\end{equation}
using the same notation as above.
Figure~\ref{fig:ScatterExplained} shows an example of a correlation plot including a graphical illustration of $|\Delta E_i^{\mathrm{reference}} - \Delta E_i^{\mathrm{benchmarked}}|$.

\section{Results}

\subsection{Energy hierarchies}

\begin{figure}
  \includegraphics[width=1\textwidth]{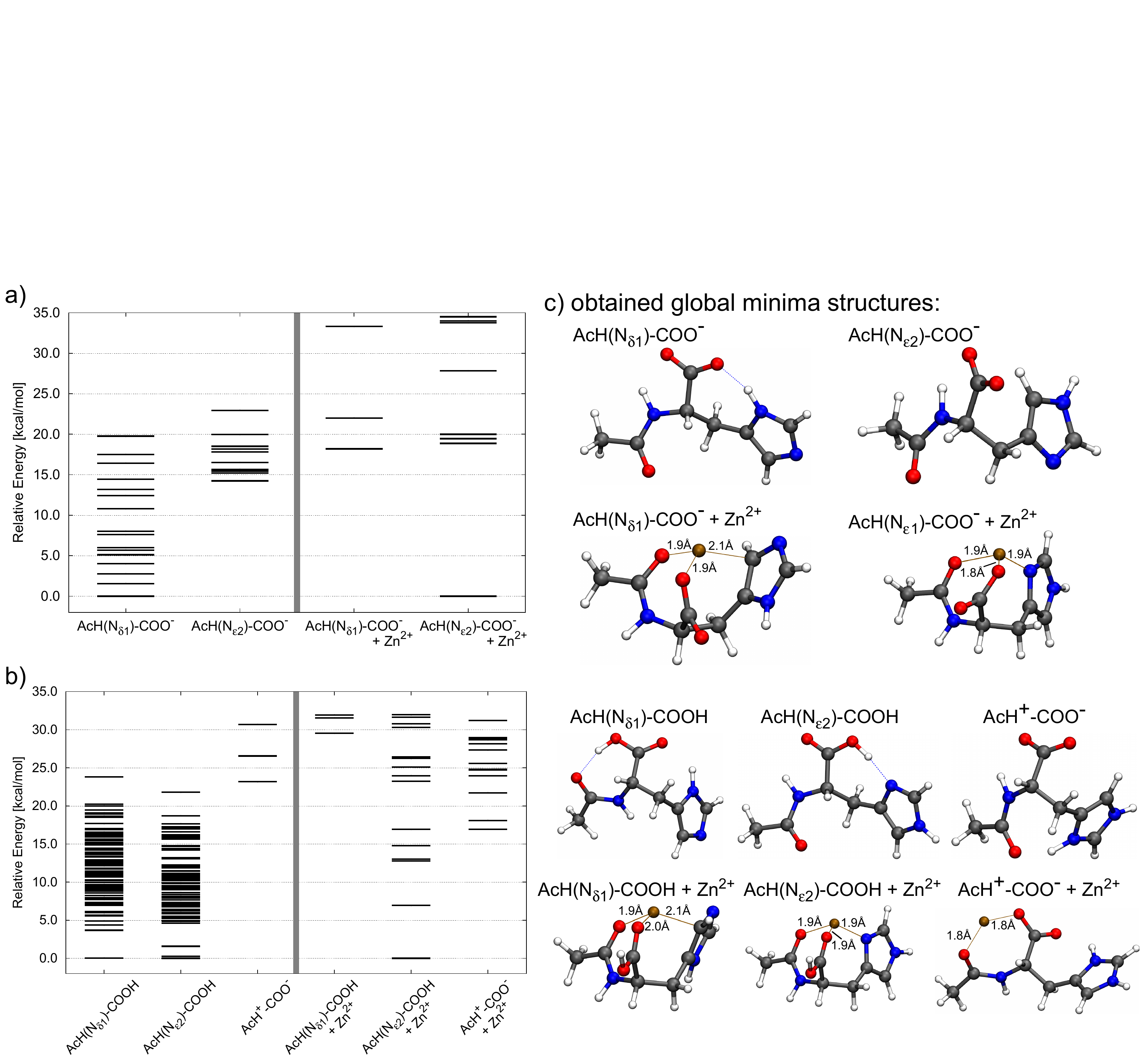}%
  \caption{\label{fig:Hierarchies} Obtained energy hierarchies at the PBE0+MBD level after having completed the conformational search for (a)~negatively charged AcH, bare and with an additional \ce{Zn^{2+}}, and (b)~neutral AcH, bare and with an additional \ce{Zn^{2+}}. (c)~For each depicted protonation state, the geometrical structure of the lowest-energy conformer is illustrated.}
\end{figure}

Figure~\ref{fig:Hierarchies} shows the obtained energy hierarchies at the PBE0+MBD level after having completed the conformational search for each individual protonation state of bare negatively charged AcH and bare neutral AcH, as well as both systems in presence of a \ce{Zn^{2+}} cation.

Comparing the two protonation states for negatively charged AcH, \textit{i.e.} \ce{AcH(N_{$\delta$1})-COO-} and \ce{AcH(N_{$\epsilon$2})-COO-} (see Figure~\ref{fig:Hierarchies}(a)), it is immediately evident that the protonation of the nitrogen atoms of the imidazole ring has a large impact concerning energy and structure of the system.
The lowest-energy conformer for \ce{AcH(N_{$\delta$1})-COO-} lies $14.3\,\mathrm{kcal/mol}$ lower in energy than the lowest-energy conformer for \ce{AcH(N_{$\epsilon$2})-COO-}, meaning that the tautomeric state of having the \ce{N_{$\delta$1}} nitrogen atom of the imodazole ring protonated is energetically much more favorable that having the \ce{N_{$\epsilon$2}} nitrogen atom protonated.
The reason for that comes abundantly clear when comparing the two lowest-energy conformers: In the case of \ce{AcH(N_{$\delta$1})-COO-}, there exists the geometrical possibility of forming a hydrogen bond between one oxygen of the carboxylate anion at the C-terminus and the nitrogen-bound hydrogen.
In case of having the \ce{N_{$\epsilon$2}} nitrogen atom protonated, a hydrogen bond cannot be formed as the proton ``points away'' from the carboxylate anion, explaining the much higher energy of this structure in comparison with its tautomeric counterpart.

The situation however changes drastically when introducing a \ce{Zn^{2+}} cation to the system. As seen in Figure~\ref{fig:Hierarchies}(a), the lowest-energy conformer of \ce{AcH(N_{$\delta$1})-COO- + Zn^{2+}} is now $18.2\,\mathrm{kcal/mol}$ higher in energy than the lowest-energy conformer of \ce{AcH(N_{$\epsilon$2})-COO- + Zn^{2+}}.
The structures look fairly similar in part as the oxygen atom of the carbonyl group at the acetylated N-terminus as well as one oxygen of the carboxylate anion are coordinated towards the \ce{Zn^{2+}}.
They differ however in the different orientation of the imidazole ring towards the cation.
In the case of \ce{AcH(N_{$\epsilon$2})-COO- + Zn^{2+}}, the deprotonated \ce{N_{$\delta$1}} atom allows for a coordinate bonding interaction with the \ce{Zn^{2+}} cation, resulting in an energetically more favorable structure compared to \ce{AcH(N_{$\delta$1})-COO- + Zn^{2+}} where the deprotonated \ce{N_{$\epsilon$2}} atom points away from the cation, resulting in an energetically less favorable cation-$\mathrm{\pi}$ interaction between imidazole ring and cation.
Adding a \ce{Zn^{2+}} cation to the system also results in an increased energetic gap between conformers.
For example, the two lowest-energy conformers of \ce{AcH(N_{$\delta$1})-COO-} are separated by $1.6\,\mathrm{kcal/mol}$ while the gap increases to $3.8\,\mathrm{kcal/mol}$ for \ce{AcH(N_{$\delta$1})-COO- + Zn^{2+}}.
For \ce{AcH(N_{$\epsilon$2})-COO-}, the two lowest-energy conformers are separated by $1.0\,\mathrm{kcal/mol}$, while the gap increases to $18.9\,\mathrm{kcal/mol}$ for \ce{AcH(N_{$\delta$1})-COO- + Zn^{2+}}.

The hierarchies of the three different protonation states of bare neutral AcH are shown in Figure~\ref{fig:Hierarchies}(b).
The global-minimum conformers of the systems of \ce{AcH(N_{$\delta$1})-COOH} and \ce{AcH(N_{$\epsilon$2})-COOH} are very similar in energy, differing only by $0.04\,\mathrm{kcal/mol}$.
For \ce{AcH(N_{$\epsilon$2})-COOH}, a hydrogen bond is possible between the deprotonated \ce{N_{$\delta$1}} atom and said proton, resulting in a very similar structure compared to system \ce{AcH(N_{$\delta$1})-COO-}.
For \ce{AcH(N_{$\delta$1})-COOH}, due to the protonated \ce{N_{$\delta$1}} atom, the proton at the carboxyl group points away from the imidazole ring and is coordinated towards the N-terminus, forming a hydrogen bond with the carbonyl group.
A protonated imidazole ring, as seen in the protonation state of system \ce{AcH+-COO-}, results in an energetically unfavorable structure, being $23.2\,\mathrm{kcal/mol}$ higher in energy than the global minimum of system \ce{AcH(N_{$\epsilon$2})-COOH}.

The situation changes again when introducing a \ce{Zn^{2+}} cation to the system, see Figure~\ref{fig:Hierarchies}(b).
The system of \ce{AcH(N_{$\epsilon$2})-COOH + Zn^{2+}} is energetically most favorable as the structure of the global minimum is very similar to the one of the system of \ce{AcH(N_{$\epsilon$2})-COO- + Zn^{2+}}: The deprotonated \ce{N_{$\delta$1}} atom allows for a coordinate bonding interaction with the \ce{Zn^{2+}} cation that in turn is also coordinated towards the electronegative oxygen atoms at the carboxyl group at the C-terminus and the carbonyl group at the N-terminus.
The global minimum of \ce{AcH+-COO- + Zn^{2+}} is $16.9\,\mathrm{kcal/mol}$ higher in energy than the global minimum of \ce{AcH(N_{$\epsilon$2})-COOH + Zn^{2+}}.
The positively charged cation and the protonated imidazole ring share no proximity, resulting in a lowest-energy structure where the \ce{Zn^{2-}} is coordinated between the oxygen of the carbonyl group at the N-terminus and one oxygen of the carboxyl group at the C-terminus.
The structure of the global minimum for \ce{AcH(N_{$\delta$1})-COO- + Zn^{2+}} is very similar to the one for \ce{AcH(N_{$\epsilon$2})-COO- + Zn^{2+}}, safe the twisted imidazole ring due to the protonated \ce{N_{$\delta$1}} atom.
Similarly to system \ce{AcH(N_{$\delta$1})-COO- + Zn^{2+}}, this results in an energetically less favorable cation-$\mathrm{\pi}$ interaction between imidazole ring and cation as the global minimum is $29.5\,\mathrm{kcal/mol}$ higher in energy than the global minimum of \ce{AcH(N_{$\epsilon$2})-COO- + Zn^{2+}}.

\subsection{Selection of minima structures}
For every protonation state, we selected the lowest-energy structures from the previous global minimum search based on energy criteria.
For one, this ensures an emphasis on the most likely structures also seen in experiment as there will always be a bias towards structures with low energy, ignoring individual set-ups or experimental conditions.
However, benchmark calculations were done including all possible protonation states for a given overall system charge, except for the case of FFs as explained above.
The large energetic differences between global minima (and consequently other low-energy conformers) of individual protonation states, as seen in Figure~\ref{fig:Hierarchies}, therefore provides a challenging benchmark testing situation for the different methods.
Table~\ref{tab:Selection} summarizes the different energy selection criteria across the systems and protonation states tackled in this work.

\begin{table}
 \caption{Minima selection criteria}
  \label{tab:Selection} 
  \begin{tabular}{l|crc}
    \hline
    \textbf{System}                          & \textbf{Net charge} & \textbf{Energy cut-off}                    & \textbf{\# Minima} \\ 
    \hline                                  
    \ce{AcH(N_{$\delta$1})-COO-}             & \multirow{2}{*}{$-1$}        & \multirow{2}{*}{$23.0\,\mathrm{kcal/mol}$} & $17$*              \\
    \ce{AcH(N_{$\epsilon$2})-COO-}           &                              &                                            & $10$*              \\
    \hline                                                                                                                                         
    \ce{AcH(N_{$\delta$1})-COO- + Zn^{2+}}   & \multirow{2}{*}{$+1$}        & \multirow{2}{*}{$41.5\,\mathrm{kcal/mol}$} & $9$                \\
    \ce{AcH(N_{$\epsilon$2})-COO- + Zn^{2+}} &                              &                                            & $9$                \\
    \hline\hline
    \ce{AcH(N_{$\delta$1})-COOH}             & \multirow{3}{*}{$0$}         & \multirow{2}{*}{$7.0\,\mathrm{kcal/mol}$}  & $11$               \\
    \ce{AcH(N_{$\epsilon$2})-COOH}           &                              &                                            & $18$               \\
    \ce{AcH+-COO-}                           &                              & $50.0\,\mathrm{kcal/mol}$                  & $8$*               \\
    \hline
    \ce{AcH(N_{$\delta$1})-COOH + Zn^{2+}}   & \multirow{3}{*}{$+2$}        & \multirow{3}{*}{$46.0\,\mathrm{kcal/mol}$} & $9$                \\
    \ce{AcH(N_{$\epsilon$2})-COOH + Zn^{2+}} &                              &                                            & $18$               \\
    \ce{AcH+-COO- + Zn^{2+}}                 &                              &                                            & $22$               \\
    \hline
    \multicolumn{4}{p{\linewidth}}{
    The ``energy cut-off'' refers to the energy relative to the respective global minimum for a given total system charge (\textit{i.e.} taking into account all possible protonation states, compare Figure~\ref{fig:Hierarchies}), within which all found minima are taken into account. The last column denotes the number of minima used for benchmarking. Numbers denoted with an asterisk (*) mean all found minima for this particular protonation state are considered.} \\
  \end{tabular}
\end{table}

\subsection{Validation of DLPNO-CCSD(T) as the reference method}

\begin{figure}
  \includegraphics[width=0.82\textwidth]{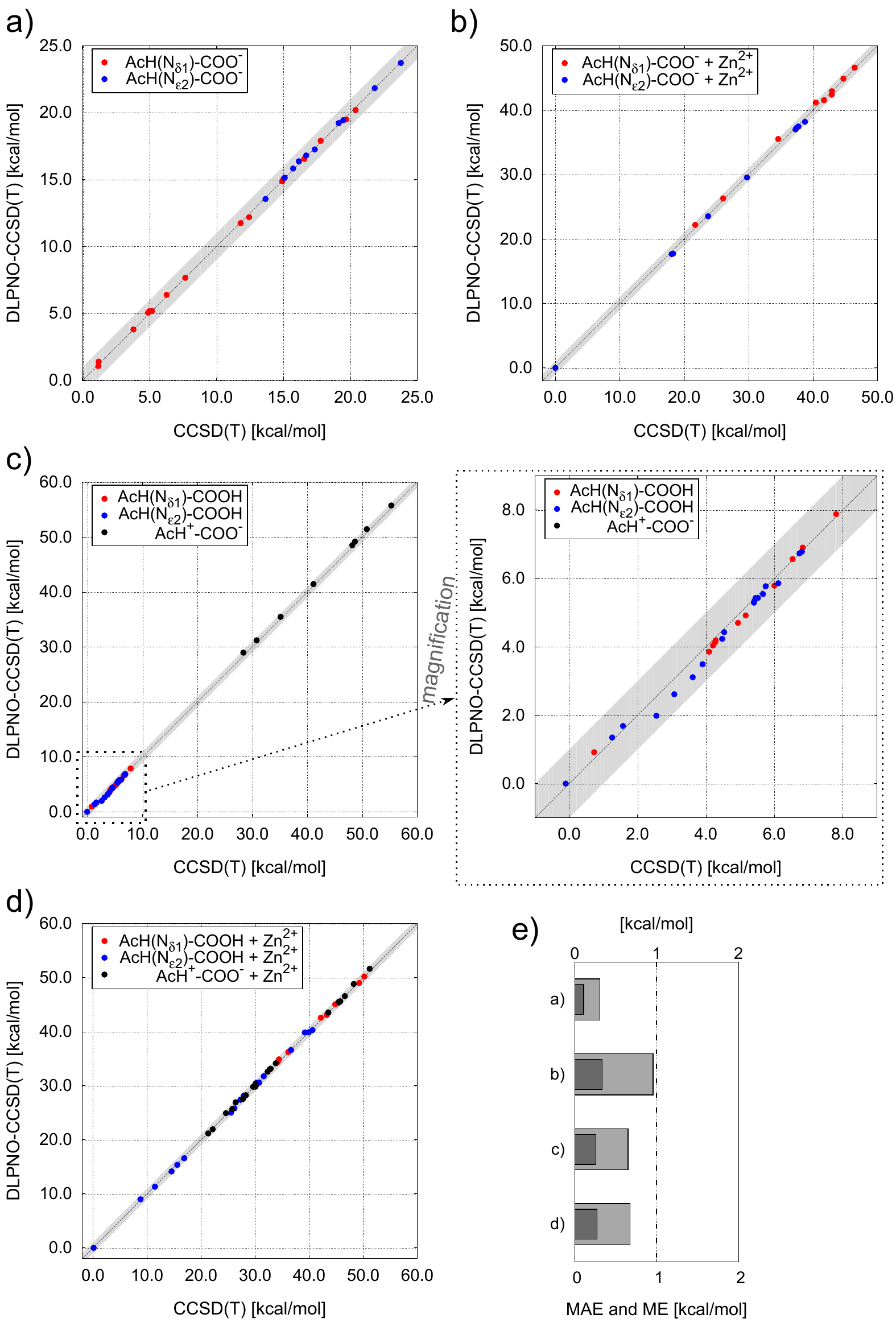}%
  \caption{\label{fig:Scatter-CCSD-DLPNO} Correlation plots for benchmarking DLPNO-CCSD(T) against conventional CCSD(T) using the ZORA-def2-SVP basis set. The systems tackled refer to (a)~negatively charged AcH, (b)~the same protonation states in presence of a \ce{Zn^{2+}} cation, (c)~bare neutral AcH, and (d)~the same protonation states in presence of a \ce{Zn^{2+}} cation. The gray shading denotes an absolute energy deviation of $1\,\mathrm{kcal/mol}$, \textit{i.e.} the region of ``chemical accuracy''. (e)~Obtained MAEs (dark-gray) and MEs (light-gray) for the four systems, following Equations~(\ref{equ:MAE}) and~(\ref{equ:ME}), respectively.}
\end{figure}

As described above and in order to validate DLPNO-CCSD(T) as the reference method used in this work, we checked the consistency of the method against conventional CCSD(T), commonly referred to as the ``golden standard of quantum chemistry''.
Calculations were laid out using Ahlrichs' relativistically recontracted ZORA-def2-SVP basis set for which CCSD(T) calculations are still affordable with respect to computational costs.
Consequently, no extrapolation or counterpoise correction had been applied here, as the intent is to compare the ``pure'' total energetic performances of both methods, which -- if similar -- will justify applying DLPNO-CCSD(T) ``instead of'' conventional CCSD(T) using larger basis sets to benchmark the other computational methods.
Figures~\ref{fig:Scatter-CCSD-DLPNO}(a)-(d) show the corresponding correlation plots for all systems tackled in this work. 
The alignment of the points near the dashed diagonal line indicates a very similar energy description between the two methods across all systems and protonation states.
To quantify that, MAEs and MEs have been computed according to Equations~(\ref{equ:MAE}) and~(\ref{equ:ME}), respectively.
For the four different systems, MAEs and MEs are given in Figure~\ref{fig:Scatter-CCSD-DLPNO}(e).
In all cases, MAEs are well within ``chemical accuracy'', \textit{i.e.} smaller than $0.5\,\mathrm{kcal/mol}$.
Furthermore, MEs are also smaller than $1\,\mathrm{kcal/mol}$ for all systems.
Taking into consideration that different protonation states and minima that differ in energy by up to more than $50\,\mathrm{kcal/mol}$ have been used, we conclude that DLPNO-CCSD(T) serves as a valid reference method for the benchmarking process of other computational methods.

In order to finally yield accurate total energies serving as benchmarks, counterpoise correction has been applied following Equation~(\ref{equ:BSSE}) and extrapolation to the complete basis set limit has been done following Equations~(\ref{equ:CBSHF}) and~(\ref{equ:CBScorr}) using Ahlrichs' relativistically recontracted ZORA-def2-SVP, ZORA-def2-TZVPP, and ZORA-def2-QZVPP basis sets.

\subsection{Benchmarking force fields and semi-empirical methods}

\begin{figure}
  \includegraphics[width=0.83\textwidth]{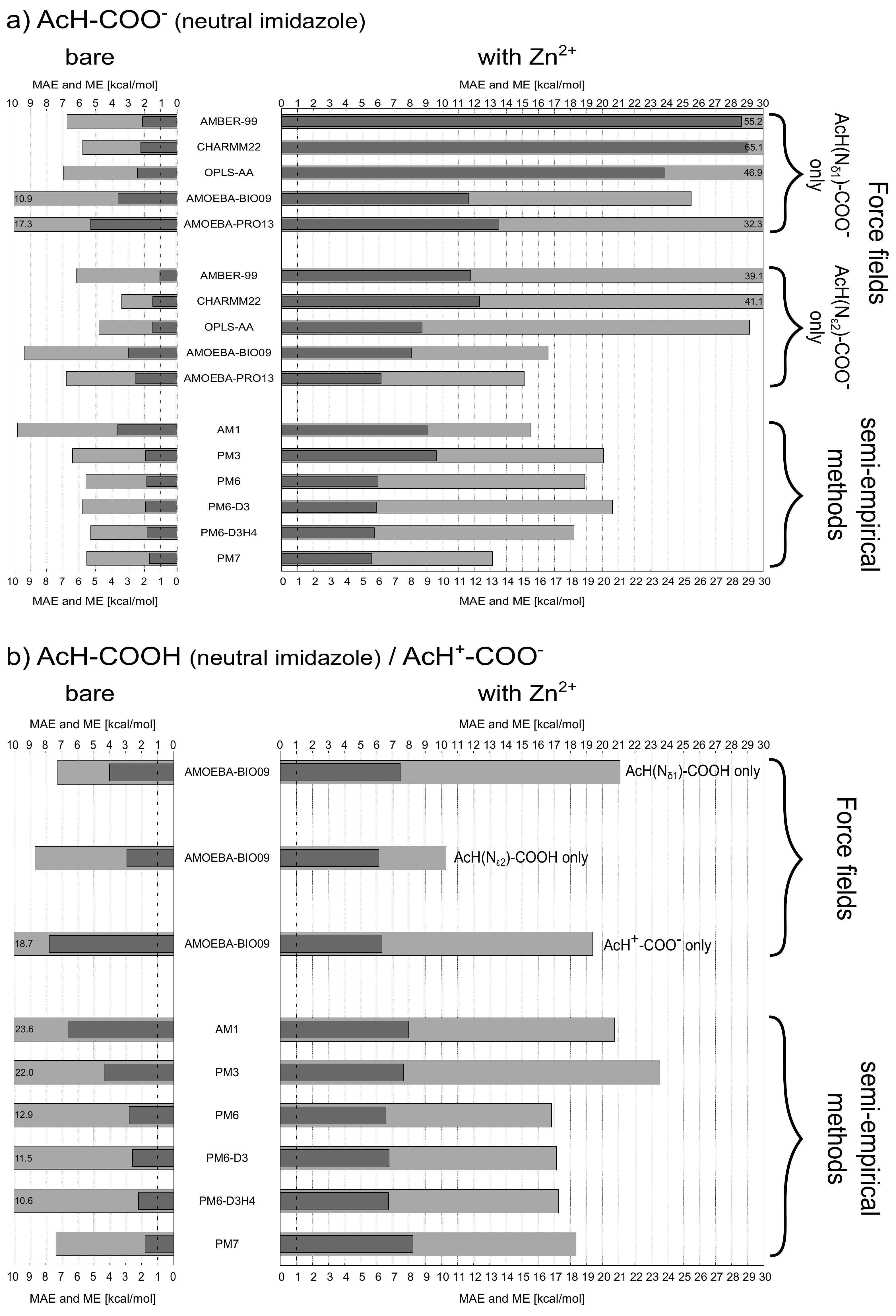}%
  \caption{\label{fig:HistosFF} MAEs (dark-gray) and MEs (light-gray) following Equations~(\ref{equ:MAE}) and~(\ref{equ:ME}) for different force fields and semi-empirical methods with respect to DLPNO-CCSD(T), for which counterpoise correction has been done following Equation~(\ref{equ:BSSE}) and extrapolation to the complete basis set limit has been done following Equations~(\ref{equ:CBSHF}) and~(\ref{equ:CBScorr}). The tackled systems are (a)~negatively charged AcH with and without a \ce{Zn^{2+}} cation, and (b)~neutral AcH with and without a \ce{Zn^{2+}}. For force fields, the different protonation states have to be treated separately.}
\end{figure}

Figure~\ref{fig:HistosFF} shows obtained MAEs and MEs calculated according to Equations~(\ref{equ:MAE}) and~(\ref{equ:ME}) for all systems tackled in this work.
As explained above, FF performance evaluation has been treated individually for different protonation states.

Considering bare neutral AcH, see Figure~\ref{fig:HistosFF}(a), conventional FFs that make use of fixed point charges are comparable in performance: For \ce{AcH(N_{$\delta$1})-COO-}, MAEs for AMBER-99, CHARMM22, and OPLS-AA have been found to be $2.1\,\mathrm{kcal/mol}$, $2.2\,\mathrm{kcal/mol}$, and $2.4\,\mathrm{kcal/mol}$, respectively.
Considering the fact that FF parameters have been derived from systems in solvation instead of gas-phase calculations applied here, the result can be considered satisfactory.
However, large MEs with up to $6.9\,\mathrm{kcal/mol}$ for OPLS-AA, indicate a possible large deviation in the energetic description for individual conformers.
Somehow surprisingly, polarizable atomic multipole-based FFs AMOEBA-BIO09 and AMOEBA-PRO13 perform worse than their FF counterparts using fixed point charges.
Large MEs up to $10.9\,\mathrm{kcal/mol}$ and $17.3\,\mathrm{kcal/mol}$ for AMOEBA-BIO09 and AMOEBA-PRO13, respectively, indicate severe discrepancies in the energetic description for individual conformers.
Consequently, the corresponding MAEs of $3.6\,\mathrm{kcal/mol}$ and $5.3\,\mathrm{kcal/mol}$ are larger than for conventional FFs.

Qualitative similar results are found for \ce{AcH(N_{$\epsilon$2})-COO-}.
Best performance for FFs is found using CHARMM22 with a MAE of $1.5\,\mathrm{kcal/mol}$ and a ME of $3.3\,\mathrm{kcal/mol}$.

Semi-empirical methods show a comparable performance to FFs, but are able to describe both protonation states simultaneously, per definition.
Best performance is found for PM7 with a MAE of $1.7\,\mathrm{kcal/mol}$ and a ME of $5.5\,\mathrm{kcal/mol}$.
For PM6, adding a long-range dispersion treatment method, \textit{i.e.} D3 or D3H4, yields very similar results of approximately $1.9\,\mathrm{kcal/mol}$, as is expected for a system of such small size.

Solvating the system with a single \ce{Zn^{2+}} cation results in very poor performances for both FFs and semi-empirical methods.
Out of the conventional FFs, OPLS-AA shows the best performance with a still very large MAE of $23.8\,\mathrm{kcal/mol}$ for \ce{AcH(N_{$\delta$1})-COO- + Zn^{2+}} and $8.7\,\mathrm{kcal/mol}$ for \ce{AcH(N_{$\epsilon$2})-COO- + Zn^{2+}}.
Polarizable atomic multipole-based FFs perform slightly better with a MAE of $11.7\,\mathrm{kcal/mol}$ using AMOEBA-BIO09 for \ce{AcH(N_{$\delta$1})-COO- + Zn^{2+}} and a MAE of $6.2\,\mathrm{kcal/mol}$ using AMOEBA-PRO13 for \ce{AcH(N_{$\epsilon$2})-COO- + Zn^{2+}}.
Semi-empirical methods show a further improvement, with PM7 yielding a MAE of $5.6\,\mathrm{kcal/mol}$ and a ME of $13.1\,\mathrm{kcal/mol}$.

As AMOEBA-BIO09 was the only FF available providing parameters out-of-the-box for the neutral carboxyl group (\ce{-COOH}), FF calculations for systems containing neutral AcH were only laid out using this particular FF, see Figure~\ref{fig:HistosFF}(b).
Protonation states with a neutral imidazole ring yield a MAE of $4.0\,\mathrm{kcal/mol}$ for \ce{AcH(N_{$\delta$1})-COOH} and $2.9\,\mathrm{kcal/mol}$ for \ce{AcH(N_{$\epsilon$2})-COOH}.
For \ce{AcH+-COO-}, performance is again very poor yielding a MAE of $7.8\,\mathrm{kcal/mol}$ and a ME of $18.4\,\mathrm{kcal/mol}$.
When adding a \ce{Zn^{2+}} cation to the system, the MAE for AMOEBA-BIO09 is larger than $6\,\mathrm{kcal/mol}$ for all three protonation states.
Out of the semi-empirical methods, PM6 performs best with a MAE of $6.6\,\mathrm{kcal/mol}$.

\subsection{Benchmarking standard DFAs and methods beyond}

\begin{figure}
  \includegraphics[width=1\textwidth]{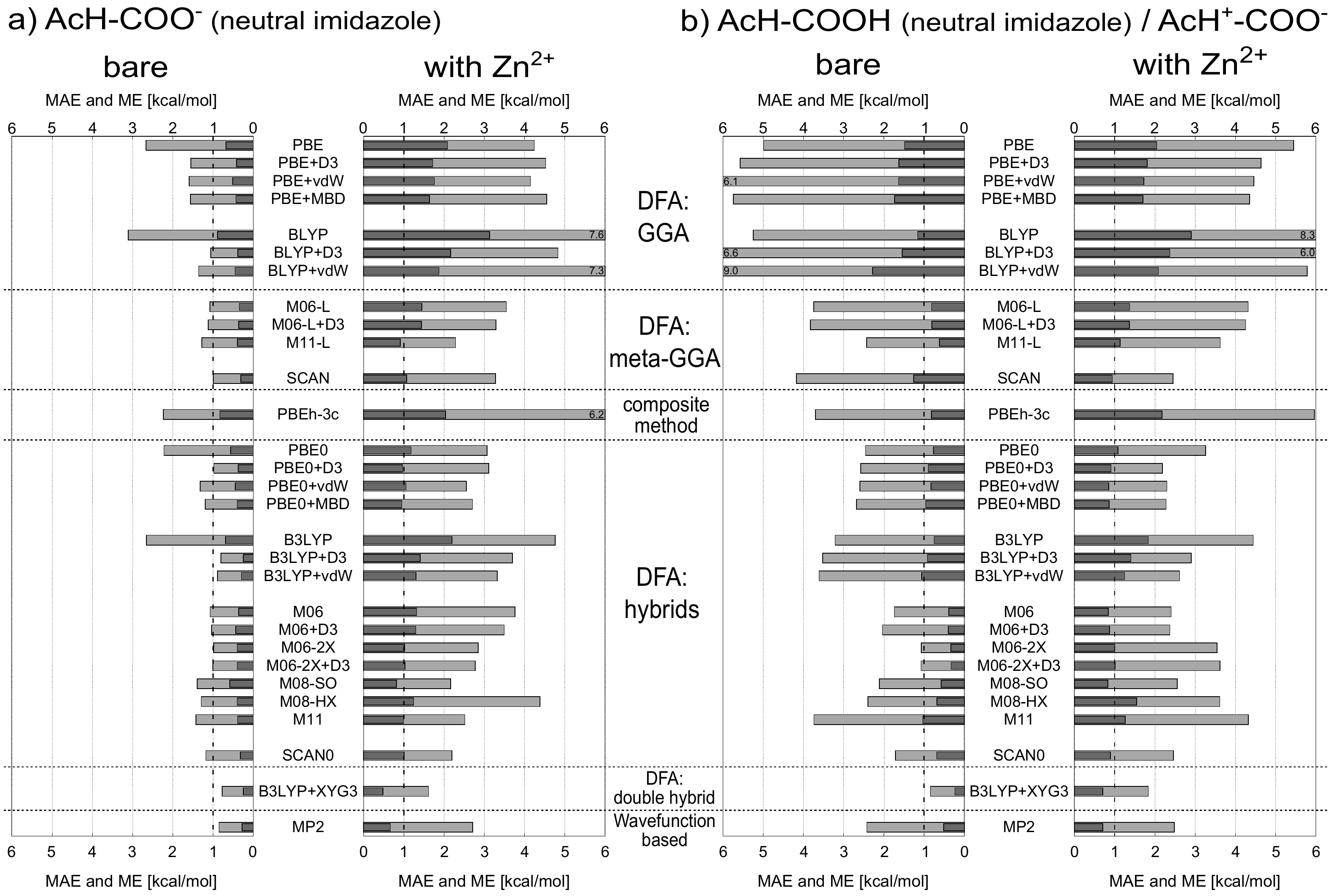}%
  \caption{\label{fig:HistosDFT} MAEs (dark-gray) and MEs (light-gray) following Equations~(\ref{equ:MAE}) and~(\ref{equ:ME}) for different standard DFAs, the composite method PBEh-3c, double hybrid DFA B3LYP+XYG3, and the wavefunction-based MP2 method with respect to DLPNO-CCSD(T), for which counterpoise correction has been done following Equation~(\ref{equ:BSSE}) and extrapolation to the complete basis set limit has been done following Equations~(\ref{equ:CBSHF}) and~(\ref{equ:CBScorr}). The tackled systems are (a)~negatively charged AcH with and without a \ce{Zn^{2+}} cation, and (b)~neutral AcH with and without a \ce{Zn^{2+}}.}
\end{figure}

Similarly to the previous section, the benchmarking process has been repeated for different kinds of DFAs as well as the wavefunction-based MP2 method.
Figure~\ref{fig:HistosDFT} shows obtained MAEs and MEs calculated according to Equations~(\ref{equ:MAE}) and~(\ref{equ:ME}) for all systems tackled in this work.
Considering bare neutral AcH, see Figure~\ref{fig:HistosDFT}(a), it is interesting to note that all tested methods already provide a very good accuracy as the MAE is less than $1\,\mathrm{kcal/mol}$ in all cases.
Out of the applied GGA xc functionals, BLYP+D3 shows best performance with a MAE of $0.4\,\mathrm{kcal/mol}$ and a ME of $1.1\,\mathrm{kcal/mol}$.
It is interesting to see that the applied long-range dispersion schemes all show significant improvement over the methods without such treatment already for systems of such a small size, compare \textit{e.g.} the ME of $3.1\,\mathrm{kcal/mol}$ for BLYP with the obtained ME of $1.1\,\mathrm{kcal/mol}$ for BLYP+D3.
All three different van der Waals treatment methods show a similar performance as the respective obtained MAEs differ by less than $0.1\,\mathrm{kcal/mol}$.
Out of the meta-GGA xc functionals, SCAN performs best with a MAE of $0.3\,\mathrm{kcal/mol}$ and a ME of $1.0\,\mathrm{kcal/mol}$.
Performance of the composite method PBEh-3c is comparable to the bare hybrid xc functional PBE0 with a MAE of $0.8\,\mathrm{kcal/mol}$ and a ME of $2.2\,\mathrm{kcal/mol}$.
Again, long-range dispersion treatments applied \textit{a posteriori} to the hybrid xc functional calculations improve the performance significantly, compare \textit{e.g.} the ME of $2.7\,\mathrm{kcal/mol}$ for B3LYP with the obtained ME of $0.8\,\mathrm{kcal/mol}$ for B3LYP+D3.
The double hybrid xc functional B3LYP+XYG3 and the wavefunction-based MP2 method perform equally well with a ME of $0.8\,\mathrm{kcal/mol}$ and $0.9\,\mathrm{kcal/mol}$, respectively.

When adding a \ce{Zn^{2+}} cation to the system, GGA xc functionals are no longer able to describe the energies within ``chemical accuracy''.
Best performance is found for PBE+MBD with a MAE of $1.6\,\mathrm{kcal/mol}$ and a ME of $4.6\,\mathrm{kcal/mol}$.
Meta-GGA xc functionals already yield a big improvement as the M11-L xc functional yields a MAE of $0.9\,\mathrm{kcal/mol}$.
The composite method PBEh-3c is not sufficient to describe energies of such systems accurately enough as the MAE is found to be $2.0\,\mathrm{kcal/mol}$ and a rather large ME of $6.2\,\mathrm{kcal/mol}$ is obtained.
Hybrid xc functionals provide a generally more accurate energetic description as PBE0+D3, PBE0+MBD, M06-2X, M06-2X+D3, M08-SO, M11 , and SCAN0 yield MAEs within $1.0\,\mathrm{kcal/mol}$.
MP2 yields a MAE of $0.7\,\mathrm{kcal/mol}$.
Out of all methods, the double hybrid xc functional B3LYP+XYG3 performs best with a MAE of $0.5\,\mathrm{kcal/mol}$ and a ME of $1.6\,\mathrm{kcal/mol}$.

For neutral bare AcH, see Figure~\ref{fig:HistosDFT}(b), the benchmarking process is much more challenging as three different protonation states are considered, as well as minima that differ in energy by up to more than $50\,\mathrm{kcal/mol}$.
Hence, GGA xc functionals are not able to yield MAEs within ``chemical accuracy''.
Best performance is seen for BLYP with a MAE of $1.2\,\mathrm{kcal/mol}$ and a ME of $5.2\,\mathrm{kcal/mol}$.
Meta-GGA xc functionals already show a big improvement with M11-L giving the best performance with a MAE of $0.6\,\mathrm{kcal/mol}$.
The composite method PBEh-3c also yields a small MAE of $0.8\,\mathrm{kcal/mol}$  while the corresponding ME of $3.7\,\mathrm{kcal/mol}$ indicates that larger energetic deviations are possible for individual conformers.
Hybrid xc functionals again perform very well as all MAEs are within $1.0\,\mathrm{kcal/mol}$.
Best performance is found for M06-2X with a MAE of $0.3\,\mathrm{kcal/mol}$ and a ME of $1.1\,\mathrm{kcal/mol}$.
Out of all methods, best performance is again found for B3LYP+XYG3 with a MAE of $0.2\,\mathrm{kcal/mol}$ and a ME of $0.8\,\mathrm{kcal/mol}$.

When adding a \ce{Zn^{2+}} cation to the system, performance of the methods is comparable to \ce{AcH+-COO- + Zn^{2+}}.
GGA xc functionals all yield a MAE above $1\,\mathrm{kcal/mol}$.
In order to reach ``chemical accuracy'' one needs to rely on meta-GGA where the SCAN functional yields a MAE of $0.9\,\mathrm{kcal/mol}$.
Out of the hybrid xc functionals, PBE0+D3, PBE0+vdW, PBE+MBD, M06, M06+D3, M06-2X, M06-2X+D3, M06-SO, and SCAN0 yield MAEs within $1.0\,\mathrm{kcal/mol}$.
Out of all methods, best performance is again found for B3LYP+XYG3 with a MAE of $0.7\,\mathrm{kcal/mol}$ and a ME of $1.8\,\mathrm{kcal/mol}$.

\subsection{Considering calculation times}

For applications, one not only needs to consider the accuracy of a particular method, but also the required computational costs and times.
All FF and semi-empirical calculations in this work have been laid out on a single CPU core and took between $0.1\,s$ and $0.3\,s$ per single-point energy evaluation.
Timings of these methods are all similar due to the small size of the benchmark systems.
Because of the fast timings of energy evaluations, conventional FFs are applied if an excessive amount of single-point energy evaluations is required, \textit{e.g.} for molecular dynamics simulations or conformational searches~\cite{HelixZn}.
However, as seen in this work where this energy description model was found only acceptable for bare neutral AcH, one should generally cross-check with other more accurate methods.

Concerning DFT calculations, timings depend on the applied xc functional, used basis set, the system, and the implementation of the method itself.
On a machine with 32 CPU cores and for the system of \ce{AcH(N_{$\delta$1})-COOH + Zn^{2+}}, it took $43\,\mathrm{s}$ on average for a single-point energy calculation with \texttt{FHI-aims} applying the GGA xc functionals PBE and BLYP, using \texttt{tier 2} basis sets and \texttt{really\_tight} settings.
Using the SCAN xc functional and the M11-L meta-GGA xc functional with the same settings took $69\,\mathrm{s}$ and $97\,\mathrm{s}$ on average, respectively. 
Calculations for the two best performant hybrid xc functionals M08-SO and SCAN0 took $839\,\mathrm{s}$ and $586\,\mathrm{s}$ on average using the same settings.

However, for most DFT production purposes one would not rely on computationally costly, yet very accurate, \texttt{really\_tight} settings, as done in this work. 
For standard cases, \texttt{tight} settings in combination with \texttt{tier 2} basis sets already provide $\mathrm{meV}$-level accurate energy differences~\cite{FHI-aims}, \textit{i.e.} within $0.02\,\mathrm{kcal/mol}$.
Indeed, repeating the procedure for the PBE, BLYP, PBE0, and B3LYP xc functionals but using \texttt{tight} settings yields virtually identical results.
On a machine with 32 CPU cores and for the system of \ce{AcH(N_{$\delta$1})-COOH + Zn^{2+}}, computational time gets then reduced from $43\,\mathrm{s}$ to $27\,\mathrm{s}$ on average for a single-point energy calculation applying the GGA xc functionals PBE.
Similarly for the hybrid xc functional PBE0, average calculation times of $738\,\mathrm{s}$ with \texttt{really\_tight} settings get reduced to $727\,\mathrm{s}$ with \texttt{tight} settings.

The composite method PBEh-3c that gave MAEs within ``chemical accuracy'' for the systems without a \ce{Zn^{2+}} cation, took $213\,\mathrm{s}$ on average for a single-point energy calculation on a single CPU core using \texttt{ORCA}.
The most accurate method across all systems and protonation states, B3LYP+XYG3, took $751\,\mathrm{s}$ on average for a single-point energy evaluation on a machine with 32 CPU cores using \texttt{FHI-aims}.
While MAEs for MP2 are comparable with B3LYP+XYG3 and within ``chemical accuracy'', energy evaluation times are much larger due to the large basis sets required for accurate predictions.
On a machine with 32 CPU cores, it took $2242\,\mathrm{s}$ on average for an MP2 energy calculation using \texttt{ORCA} and the ZORA-def2-QZVPP basis set.

\section{Conclusions}

We examined the goodness of commonly applied levels of theory, \textit{i.e.} force fields, semi-empirical methods, density-functional approximations (DFA), and wavefunction-based methods with respect to high-level coupled-cluster calculations.
To that end, benchmark systems consisting of either a bare acetylhistidine or microsolvated with a \ce{Zn^{2+}} cation were (i)~conformationally sampled by performing a global energy minimum search combining both FF and DFA, and (ii)~obtained conformational minima were used for benchmarking against DLPNO-CCSD(T) single-point energy-calculations.

For bare negatively charged AcH, the obtained energy hierarchies on the hybrid DFA level showed that the protonation state of \ce{AcH(N_{$\delta$1})-COO-} is energetically favorable compared to \ce{AcH(N_{$\epsilon$2})-COO-} as the respective global minima differ by $14.3\,\mathrm{kcal/mol}$ in energy.
The situation is reversed when adding a \ce{Zn^{2+}} cation to the system: the protonation state of \ce{AcH(N_{$\epsilon$2})-COO- + Zn^{2+}} is energetically preferred as the respective global minima differ by $18.2\,\mathrm{kcal/mol}$.
Considering bare neutral AcH, the two protonation states of \ce{AcH(N_{$\delta$1})-COOH} and \ce{AcH(N_{$\epsilon$2})-COOH} yield global minima that are similar in energy.
When adding a \ce{Zn^{2+}} cation to the system, \ce{AcH(N_{$\epsilon$2})-COO- + Zn^{2+}} is energetically preferred to the other two protonation states of \ce{AcH+-COO- + Zn^{2+}} and \ce{AcH(N_{$\delta$1})-COO- + Zn^{2+}}, as the global minima differ by $16.9\,\mathrm{kcal/mol}$ and $29.5\,\mathrm{kcal/mol}$ in energy, respectively.

The benchmarking process, based on single-point energy calculations and assessed by means of MAEs and MEs, revealed that force fields and semi-empirical methods are generally not reliable enough for an energetic description of these systems within ``chemical accuracy'' of $1\,\mathrm{kcal/mol}$.
While GGA xc functionals like PBE and BLYP, as well as the composite method PBEh-3c have problems in their energetic description for systems containing a \ce{Zn^{2+}} cation, it is possible to reach ``chemical accuracy'' for all systems already using the meta-GGA SCAN xc functional.
Hybrid xc functionals perform generally well with MAEs within $1\,\mathrm{kcal/mol}$ for most of them.
Out of the hybrid xc functionals, best performance is shown for M06-SO and SCAN0.
Out of all tested methods, the double hybrid xc functional B3LYP+XYG3 resembles the benchmark method DLPNO-CCSD(T) best with a MAE of $0.7\,\mathrm{kcal/mol}$ and a ME of $1.8\,\mathrm{kcal/mol}$.
While MP2 performs similarly as B3LYP+XYG3, computational costs, \textit{i.e.} timings, are increased by a factor of 4 in comparison due to the large basis sets required for accurate results.

\begin{acknowledgement}
The authors thank the joint Max-Planck-EPFL Center for Molecular Nanoscience and Technology for financial support and Prof. Matthias Scheffler for his continuous support.

\end{acknowledgement}

\section{Supporting Information}
\textbf{SI.zip} The \textit{zip}-archive includes a collection of all found conformers (\textit{xyz} file format) and corresponding energies for the obtained hierarchies at the PBE0+MBD level after having completed the conformational search, see Figure~\ref{fig:Hierarchies}. In addition, the \textit{zip}-archive also includes listings of all calculated single-point energies for all computational methods and benchmark structures.
See here:\\
\begin{footnotesize}
\url{http://w0.rz-berlin.mpg.de/user/baldauf/carsten_pdf/Schneider_HisBench_2018_SI.zip}
\end{footnotesize}


\providecommand{\latin}[1]{#1}
\makeatletter
\providecommand{\doi}
  {\begingroup\let\do\@makeother\dospecials
  \catcode`\{=1 \catcode`\}=2 \doi@aux}
\providecommand{\doi@aux}[1]{\endgroup\texttt{#1}}
\makeatother
\providecommand*\mcitethebibliography{\thebibliography}
\csname @ifundefined\endcsname{endmcitethebibliography}
  {\let\endmcitethebibliography\endthebibliography}{}

\end{document}